\documentclass[]{spie}  

\usepackage{amsmath,amsfonts,amssymb}
\usepackage{graphicx}
\usepackage[colorlinks=true, allcolors=blue]{hyperref}

\title{Combined visible and near-infrared OPA for wavelength scaling experiments in strong-field physics.}

\author[a]{David T. Lloyd}
\author[b]{Kevin O'Keeffe}
\author[c]{Adam S. Wyatt}
\author[a]{Patrick N. Anderson}
\author[a]{Daniel Treacher}
\author[a]{Simon M. Hooker}
\affil[a]{Department of Physics, University of Oxford, Clarendon Laboratory, Parks Road, Oxford, OX1 3PU, UK}
\affil[b]{Department of Physics, Swansea University, Singleton Park, Swansea, SA2 8PP, UK}
\affil[c]{Central Laser Facility, STFC Rutherford Appleton Laboratory, Harwell OX11 0QX, UK}

\authorinfo{Further author information: (Send correspondence to david.lloyd@physics.ox.ac.uk)}

\begin{document} 
\maketitle

\begin{abstract}
We report the operation of an optical parametric amplifier (OPA) capable of producing gigawatt peak-power laser pulses with tunable wavelength in either the visible or near-infrared spectrum. The OPA has two distinct operation modes (i) generation of $>350 \:\mu \mathrm{J}$, sub 100 fs pulses, tunable between 1250 -- 1550 nm; (ii) generation of $>190\: \mu \mathrm{J}$, sub 150 fs pulses tunable between 490 -- 530 nm. We have recorded high-order harmonic spectra over a wide range of driving wavelengths. This flexible source of femtosecond pulses presents a useful tool for exploring the wavelength-dependence of strong-field phenomena, in both the multi-photon and tunnel ionization regimes.\\ This paper was published in Proceedings of SPIE 10088, Nonlinear Frequency Generation and Conversion: Materials and Devices XVI. \url{http://dx.doi.org/10.1117/12.2250775.} Please cite as:\\
David T. Lloyd \emph{et al.}, `` Combined visible and near-infrared OPA for wavelength scaling experiments in strong-field physics,'' \emph{Proc.} \emph{SPIE} 10088, Nonlinear Frequency Generation and Conversion: Materials and Devices XVI, 1008815 (February 20 2017).\\
\textcopyright (2017) COPYRIGHT Society of Photo-Optical Instrumentation Engineers (SPIE).
\end{abstract}

\keywords{Parametric amplifier,
 Strong-field physics,
 Nonlinear frequency conversion,
 Femtosecond pulses,
 High peak power,
 High-order harmonics, High harmonic generation, Tunable wavelength 
}

\section{INTRODUCTION}

This year marks the twentieth anniversary of the publication of Wilson and Yakolev's seminal work ``Ultrafast rainbow: tunable ultrashort pulses from a solid-state kilohertz system''\cite{Wilson1997} in which the authors described and demonstrated how cascaded non-linear processes could vastly extend the wavelength coverage of ultrafast laser sources. In the intervening years large strides have been made in realising the concept of the ``Ultrafast Rainbow''. In particular, nonlinear parametric processes have been utilised to produce laser pulses with exceptional properties \cite{Fattahi2014}, often using titanium-sapphire pump lasers.

\begin{figure} [ht]
   \begin{center}
   \begin{tabular}{c} 
   \includegraphics[height=8cm]{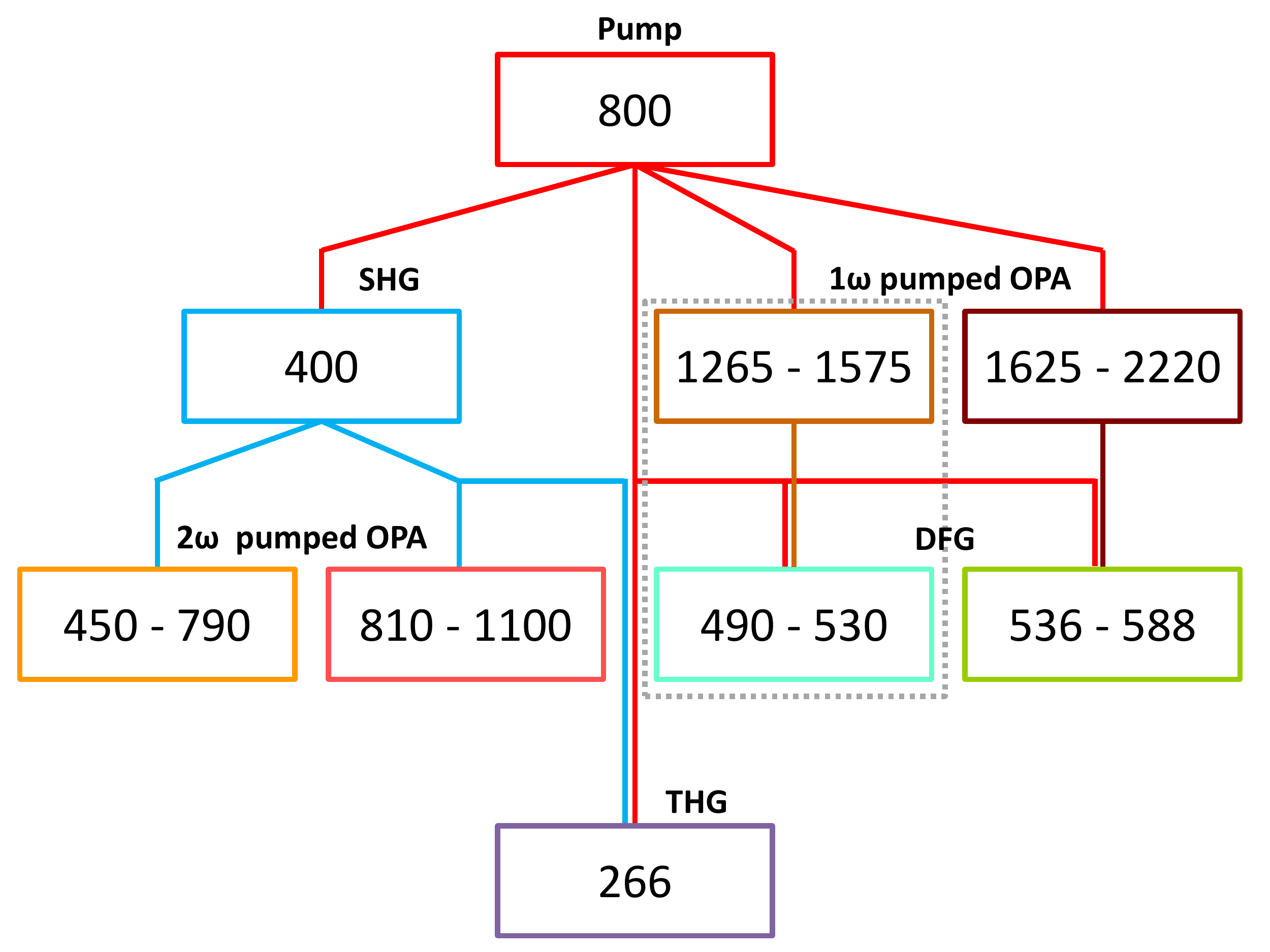}
   \end{tabular}
   \end{center}
   \caption[example] 
   { \label{Rainbow} 
Some potential paths for nonlinear wavelength conversion of a high energy 800 nm wavelength pump pulse: SHG - second harmonic generation, THG - third harmonic generation, DFG - difference frequency generation. The scheme exploited in this paper is enclosed by the grey dotted lines.}
   \end{figure}

Possible routes for wavelength conversion of a fixed wavelength pump beam are presented in fig. \ref{Rainbow}. Sum and difference-frequency generation, as well as the combination of both processes allow for spectral coverage from the ultraviolet (UV) to mid-infrared (MIR). The approach adopted in our work makes use of both parametric amplification and sum frequency generation in a cascaded scheme, enclosed by the dotted line in fig. \ref{Rainbow}.

In recent years optical parametric amplifiers have become a mainstay in ultrafast laser laboratories, in particular for the generation of high energy, few-cycle, passively carrier envelope phase stable \cite{Baltuska2002a}, mid-infrared wavelength waveforms.\cite{Nisoli1998,Zhang2009,Silva2012} The large ponderomotive potential associated with infrared pulses enables both the generation of high energy electrons, for laser-induced electron diffraction studies \cite{Wolter}, as well as the generation of kilo-electronvolt energy photons \cite{Popmintchev2012a} through the process of high harmonic generation (HHG).

Parametric amplification has also been used to produce sources of wavelength-tunable, ultrafast pulses in the visible spectral region. Cirmi \emph{et al.} demonstrated a three stage, non-collinear OPA pumped by the second harmonic of a titanium:sapphire laser.\cite{Cirmi2012} The signal beam wavelength could be tuned between $450-790$ nm. After temporal compression the pulses had an energy as high as $360\:\mu$J and pulse duration as short as 26 fs, from $6\:$mJ of pump pulse energy (prior to frequency doubling). However, the large group velocity mismatch when pumping BBO at 400 nm is a limiting factor for the conversion efficiency. Nevertheless, the output of this OPA was subsequently used to generate high order harmonics using visible pulses of various wavelengths.\cite{Lai2013} In contrast to earlier work, they found a less drastic wavelength dependence of the high harmonic intensity compared to when short wave infrared pulses were used, attributed in part, to the different ionization mechanism (nonadiabatic over quasistatic) accessible with visible wavelength pulses. 

Aside from exploring the wavelength scaling of strong-field phenomena, another possible use of a tunable source of high energy, ultrafast pulses was recently demonstrated through the work of Rothhardt and coauthors.\cite{Rothhardt2014} By using laser pulses with wavelength tunable between 760 -- 830 nm they were able to shift the 17th harmonic order to coincide with the $3\mathrm{s}3\mathrm{p}^6 4\mathrm{p}^1\mathrm{P}_1$ resonance in argon at 26.6 eV. This resulted in a greatly reduced photo-absorption in the generating medium, leading to a bright, spectrally narrow emission at the resonance wavelength. Such narrowband sources of short wavelength light hold great promise for applications ranging from lensless imaging to high precision spectroscopy.

This paper is organised as follows. In sec. \ref{OPA_specs} we outline the design of our OPA while in sec. \ref{Perform} we present results of characterization of the OPA output in both the SWIR and visible spectral regions. Section \ref{BLIND} briefly describes a procedure for maximising the peak intensity of the OPA output. A 1-D model of HHG is briefly outlined in section \ref{Model}. The experimental HHG results are presented in sections \ref{IRHHG} and \ref{VISHHG}, for the SWIR and visible outputs, respectively. Finally, in sect. \ref{Apps} we identify some possible further experimental studies in the area of HHG which could benefit from a source of high energy, wavelength tunable, ultrafast pulses.

\section{OPA DESIGN AND OPERATION}
\label{OPA_specs}
\begin{figure} [ht]
   \begin{center}
   \begin{tabular}{c} 
   \includegraphics[height=7cm]{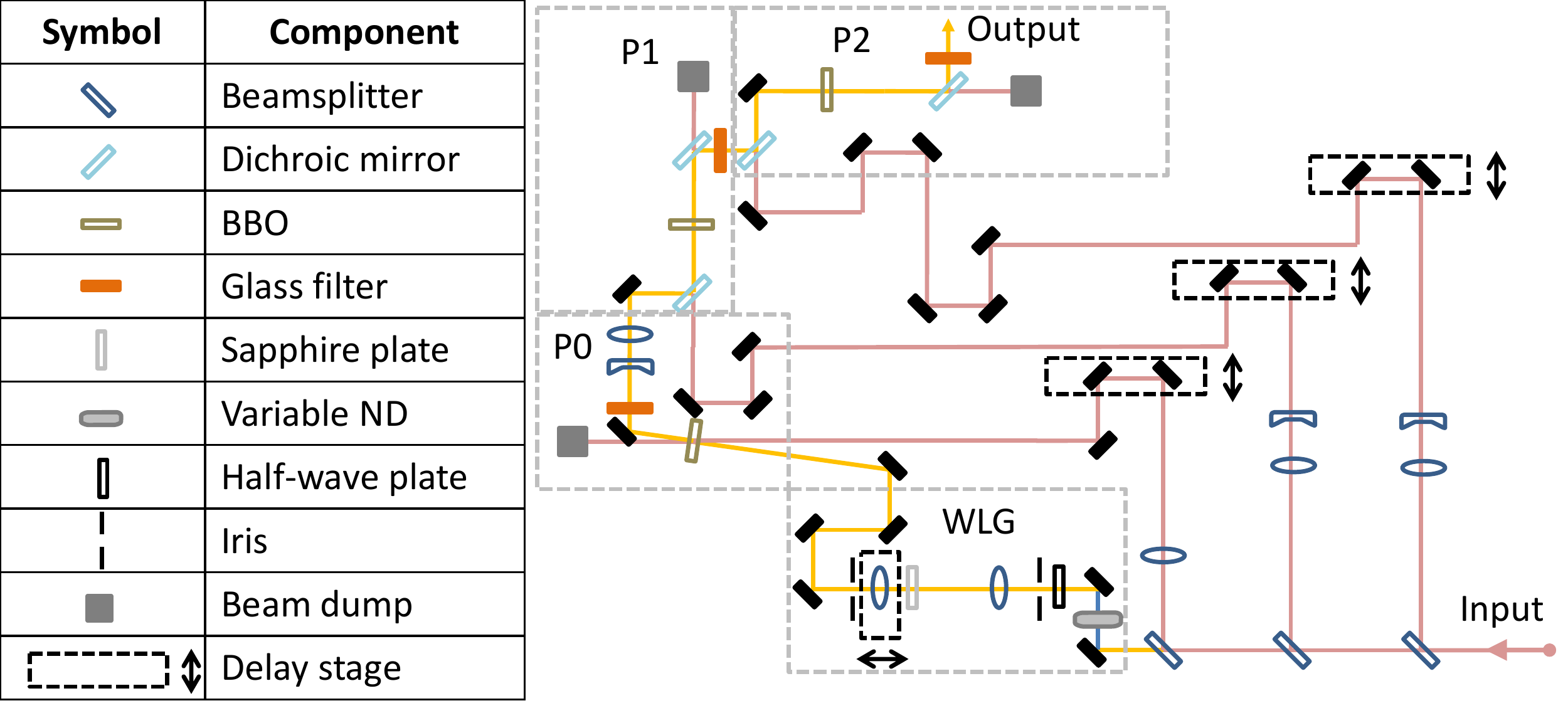}
   \end{tabular}
   \end{center}
   \caption[example] 
   { \label{Schematic} Schematic diagram of the OPA. Red represents the pump beam, yellow the seed/signal beam. The individual stages are labelled and enclosed by grey dotted lines.}
   \end{figure}

The OPA described in this article is based upon designs previously described in the literature \cite{Cerullo2003, Vozzi2007, Silva2012}. In our case, the OPA was synchronously pumped by the output of a titanium:sapphire regenerative amplifier, providing \\3 mJ, 40 fs duration pulses at a repetition rate of 1 kHz. A schematic of the OPA is shown in fig. \ref{Schematic}. The OPA has footprint of 118 by 58 cm. Throughout this paper the four distinct stages of the OPA will be identified as: white light generation (WLG), non-collinear preamplifier (P0), power amplifier (P1) and a reconfigurable final stage (P2).

\subsection{Seed Generation}
The beam in this stage was first switched to s-polarisation using an achromatic half waveplate in order to permit type II interactions in latter stages in the OPA. The seed for the OPA was generated through spectral broadening of a small fraction of the pump beam ($\approx 10 \:\mu\mathrm{J}$) focussed into a sapphire plate (2 mm thickness). An iris controlled the size of the beam prior to a $f=150\:\mathrm{mm}$ focal length lens. The intensity at the sapphire crystal was adjusted using a variable neutral density attenuator. The energy in the white light continua was typically around 1 nJ per nm bandwidth. A second iris was used to select the central part of the spectrally broadened beam, while a $f=50\:\mathrm{mm}$ achromat lens refocussed the beam into the preamplifier. 

\subsection{Non-collinear Preamplifier} 
Approximately $95\:\mu\mathrm{J}$  of the pump beam was focussed approximately 150 mm before a BBO crystal (3 mm thick, 5 mm in diameter, cut for $\theta=25.9^\circ$) using an $f=750\:\mathrm{mm}$ lens, yielding a peak intensity $\approx 250\: \mathrm{GW/cm}^{2}$. The pump beam was diverging at the crystal face to match the divergence of the white light seed and avoid self-focussing in the BBO. The white light seed and focussed pump crossed at an external angle of $\approx 2.8^\circ$. A portion of the broadband seed in the IR was parametrically amplified through type II difference frequency generation of the pump.  After the crystal an aperture separated the amplified seed from pump and idler beams, while a longpass glass filter (Thorlabs FGL1000) blocked the un-amplified, visible part of the seed spectrum. A Galilean telescope collimated the seed to an approximate diameter of 5 mm. 

\subsection{Power Amplification}

Approximately 1 mJ of the pump beam was down-collimated to a width of 3 mm, yielding a peak intensity of $\approx180$ GW/$\mathrm{cm}^2$ at the BBO crystal (4 mm thick, 7 mm in diameter, cut for $\theta = 25.9^\circ$). The pump beam and the amplified seed beam from P0 were combined prior to the BBO using a dichroic beamsplitter (Thorlabs DMSP1000), such that the two beams were collinear. After further amplification in the BBO, the signal beam was separated from the pump and idler beams using a dichroic mirror (Thorlabs DMSP1000), and further spectrally filtered using a longpass glass filter (Thorlabs FGL1000).

\subsection{Reconfigurable Final Stage}
A unique characteristic of our device is the ability to rapidly switch the output of the OPA from shortwave infrared (centre wavelength in the range 1250 -- 1550 nm) to visible (referred to as VIS - centre wavelength in the range 490 -- 530 nm) pulses.

The final stage used a 4 mm thick, 9 mm wide BBO crystal, cut at an angle of $25.9^\circ$. The phasematching angles for type II difference frequency generation are $25.8^\circ$ -- $28.3^\circ$ for the relevant wavelength range, while for type II sum frequency generation they are $27.8^\circ$ -- $30.2^\circ$, meaning that the crystal does not need to be rotated significantly in switching from one mode of operation to the other. Approximately 1.2 mJ of pump energy was down collimated to a size of 3 mm, yielding a peak intensity of 220 GW/$\mathrm{cm}^2$ at the crystal face.

For production of SWIR pulses, the crystal was used at normal incidence. The signal beam was amplified through type II parametric amplification in a single pass through the BBO crystal. A dichroic mirror (Thorlabs DMSP1000) and glass filter (Thorlabs FGL1000) separated the amplified signal beam from the pump. We will refer to this setting as ``difference-frequency mode'', after the non-linear process which allows parametric amplification to occur.

For production of VIS pulses, the crystal was rotated by 2 -- 3 degrees, with respect to the SWIR set-up\footnote{The crystal was also rotated $90^\circ$ in the azimuthal plane.}. This angle corresponds to phasematching for sum-frequency generation between the signal beam from P1 and the pump beam. After the BBO crystal, a dichroic mirror (Thorlabs DMSP550) and glass filter (Knight Optical 575FCS) separated the sum-frequency beam from the (depleted) signal and pump beams. We will refer to this setting as ``sum-frequency mode''.  We note that the function of the final two stages when the OPA is in sum-frequency mode could in principle be achieved with a single stage \cite{Petrov1995}. However the single stage approach suffers from a much reduced conversion efficiency.

Switching from difference-frequency to sum-frequency mode involves rotating the BBO and swapping the dichroic mirror and glass filters. This process was performed manually in a few minutes, but, in principle, the process could be easily automated, reducing the switch-over time.

This study is limited to the sum and difference-frequency generation between the signal beam from P1 and the pump. However, it would be possible to design a similar device which uses the idler beam from P1 thus further increasing the wavelength coverage possible. This process is shown on the right hand side of fig. \ref{Rainbow}. 
 
 \section{OPA PERFORMANCE}
\label{Perform}

\begin{figure} [ht]
   \begin{center}
   \begin{tabular}{c} 
   \includegraphics[height=15cm]{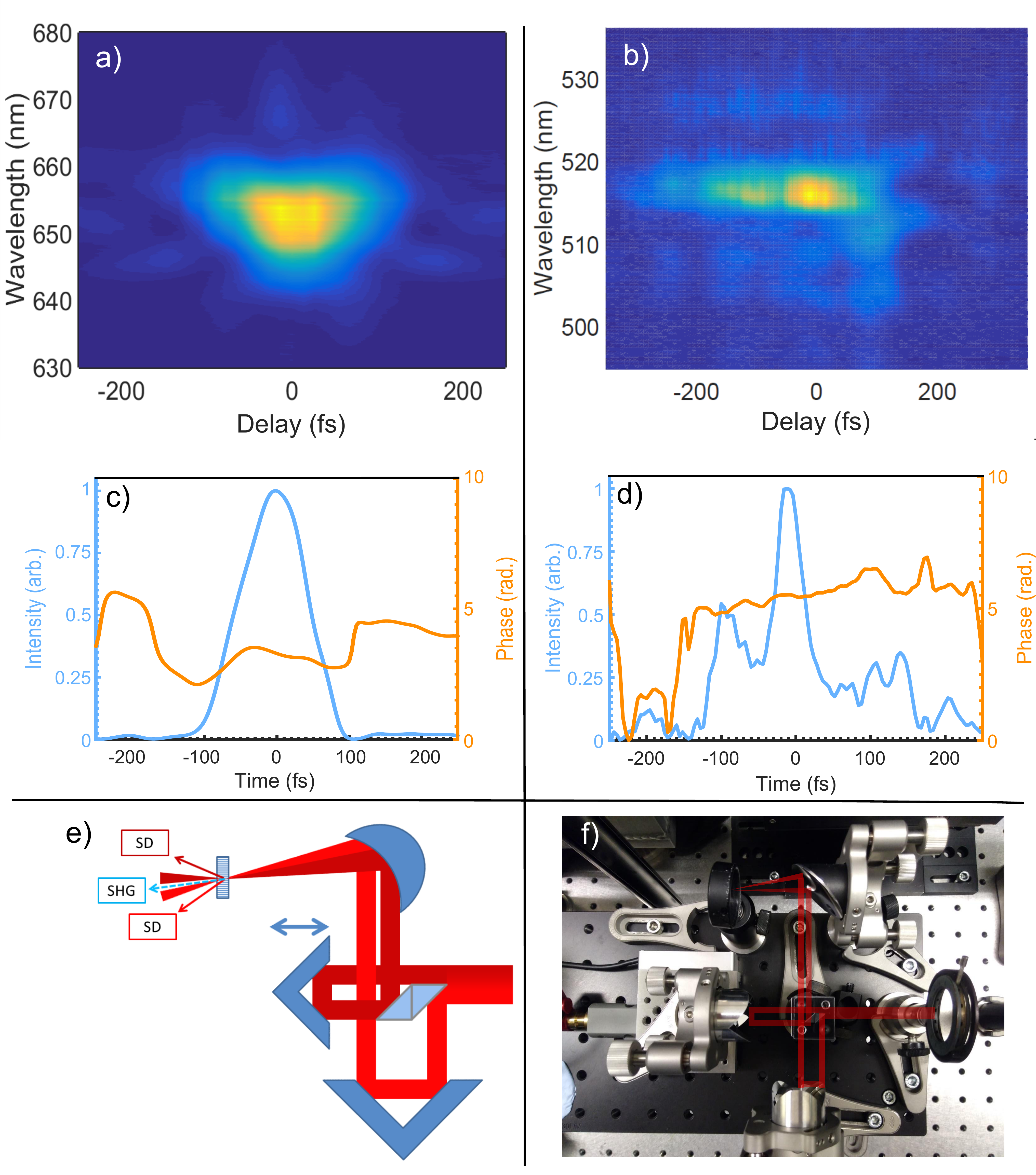}
   \end{tabular}
   \end{center}
   \caption[example] 
   { \label{FROG} 
\textbf{a)} Raw SHG FROG trace of OPA ouput in difference-frequency mode. \textbf{b)} Raw SD FROG trace of OPA output in sum-frequency mode. \textbf{c)} Pulse temporal intensity (solid blue line) and phase (solid orange line) retrieved from the trace shown in a). The FROG error was 0.007. \textbf{d)} Pulse temporal intensity (solid blue line) and phase (solid orange line) retrieved from the trace shown in b). The FROG error was 0.022. \textbf{e)} Schematic diagram of device used to perform SHG and SD FROG measurements. The two parts of the wavefront divided by the beamsplitter are shown by different shades of red. \textbf{f)} Photograph of the device used to perform SHG and SD FROG measurements, the beam path is overlaid with a red line.}
   \end{figure}

The output of the OPA in each of the two operation modes was characterized in terms of pulse energy, duration, wavelength as well as spatial mode quality. For temporal characterization a FROG device was constructed that was able to perform both second harmonic generation (SHG) and self-diffraction (SD) FROG measurements. A schematic of the device is shown in fig. \ref{FROG} e). The wavefront of the incident beam is split by an aluminium coated right angle prism. Each half of the wavefront is retroreflected by an aluminium coated hollow retroreflector, with one of the retroreflectors mounted on a closed-loop translation stage (Newport 9067-COM-E-M). The two beams are incident on an aluminium coated off-axis parabola and cross at focus in a non-linear material. In the case of the SHG FROG measurement, a 0.1 mm thick BBO crystal was used, with the second harmonic spectrum measured using a fibre-coupled spectrometer (Ocean Optics USB4000) as a function of delay. While in the case of the SD FROG measurement a 1 mm thick sapphire plate was used and the first order self-diffracted spectrum was measured as a function of delay using the same fibre coupled spectrometer. A photo of the compact arrangement is shown in fig. \ref{FROG} f).

\subsection{Difference Frequency Mode}

In difference frequency mode, at an output wavelength of $\approx 1300\:$nm, pulse energies in excess of $350\:\mu$J (signal beam only) were observed from a pump pulse energy of 3 mJ\footnote{This was the total pulse energy measured prior to the OPA.}, with lower signal pulse energies measured at other wavelengths in the range $1250 - 1550\:$nm. This corresponds to a pump energy-to-signal energy conversion efficiency of $\approx 12\%$. We note that better conversion efficiencies have been reported using both custom-built and commercial OPAs of a similar design\cite{Zhang2009, Yakovlev1994a}.

A typical SHG FROG trace of the OPA output in difference frequency mode is shown in fig. \ref{FROG} a). The reconstructed pulse, shown in fig. \ref{FROG} c) has a full width half maximum of $\approx 100\:\mathrm{fs}$, compared to $85\:\mathrm{fs}$ Fourier-limited duration. The short duration of the pulses out of the OPA are in part the result of the low amount of net chirp imparted to the signal beam in passage through the device: $|\phi_2|<500\:\mathrm{fs}^2$ over the range $\lambda =$1250--1550 nm, i.e. sufficient dispersion to stretch an 80 fs pulse by $\approx2\:$fs.

\subsection{Sum Frequency Mode}

In sum frequency mode pulse energies up to $192\:\mu$J were recorded for a pump pulse energy of $2.93\:$mJ, corresponding to a pump energy-to-output energy of $6.5\%$. We expect that further optimisation of the OPA design, including BBO length and the partition of pump energy between the stages could improve the conversion efficiency.

An SD FROG trace of the OPA output set to a wavelength of $\lambda = 517\:\mathrm{nm}$ is shown in fig. \ref{FROG} b). The reconstructed trace, shown in fig. \ref{FROG} d) shows significant structure, with the bulk of the pulse (at a $I>0.25 I_\mathrm{max}$ level) contained within $\approx 120\:$fs. Further effort is needed to explore the possibility of producing temporally ``cleaner'' visible wavelength pulses from the OPA.
 
\begin{figure} [ht]
   \begin{center}
   \begin{tabular}{c} 
   \includegraphics[height=9cm]{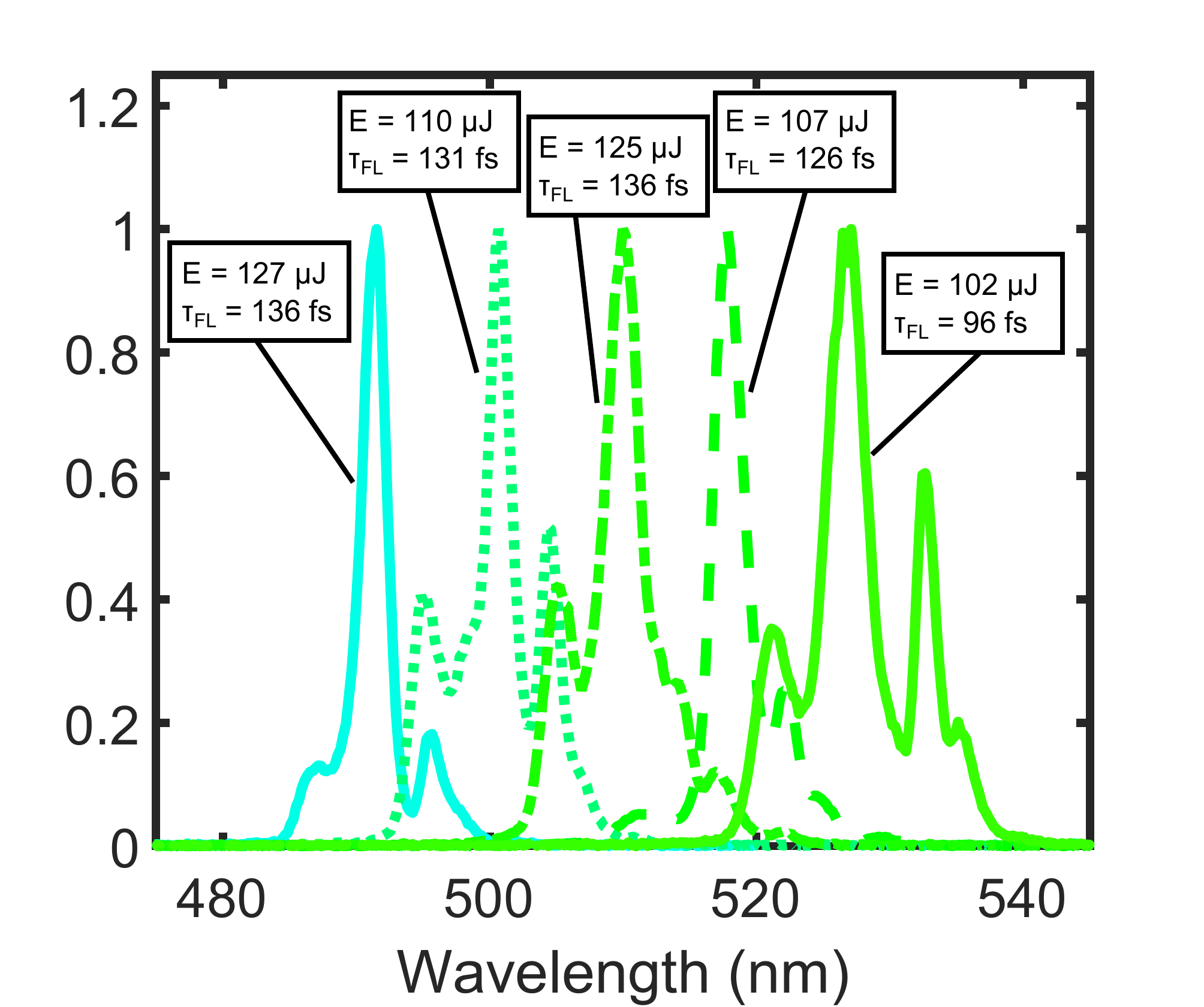}
   \end{tabular}
   \end{center}
   \caption[example] 
   { \label{GRN_Spec} 
Example spectra when the OPA was set to sum frequency mode, showing the wavelength range accessible. The colour of each line matches the central wavelength of the beam.}
   \end{figure}

In fig. \ref{GRN_Spec} spectra of the OPA output over the full VIS wavelength range are plotted. For each spectrum the pulse energy and Fourier-limited duration are also shown, demonstrating that neither quantity varies significantly with output wavelength.

\section{``BLIND'' OPTIMISATION}
\label{BLIND}
For applications in strong-field physics, the peak laser intensity is a key parameter. A high pulse energy, short pulse duration and good spatial quality are all necessary for producing the highest peak intensity possible. However, optimising the output of an OPA for any one of the three contributing parameters ($E$, $\tau$ and $w_0$) does not necessarily lead to highest peak intensity. In this section we describe a simple means of ``blind'' optimisation of the peak intensity achievable from an OPA by maximising an experimentally accessible proxy for $I_\mathrm{pk}$.

For an OPA similar to that described in section \ref{OPA_specs} there are a large number of degrees of freedom that must be adjusted to optimise the output: for each stage, up to three orthogonal angles of the crystal, the pump-seed/signal delay, and the pump chirp must be optimised. The situation is simplified by assuming that the multi-dimensional parameter space is locally convex with respect to the optimum settings. The net-result is that the output of the OPA is optimised after each of the degrees of freedom is individually optimised.

We chose to optimise the intensity of third harmonic [$I(3\omega_0)$] produced by focussing the beam in air. Since the third harmonic depends on the fundamental as $I(3\omega_0)\propto I(\omega_0)^{3/2}$, maximising the third harmonic is a means of ``blindly'' optimising the OPA peak intensity. As an added benefit, the third harmonic can be generated in situ, i.e. at the location where the strong-field experiment will take place.

\begin{figure} [ht]
   \begin{center}
   \begin{tabular}{c} 
   \includegraphics[height=12cm]{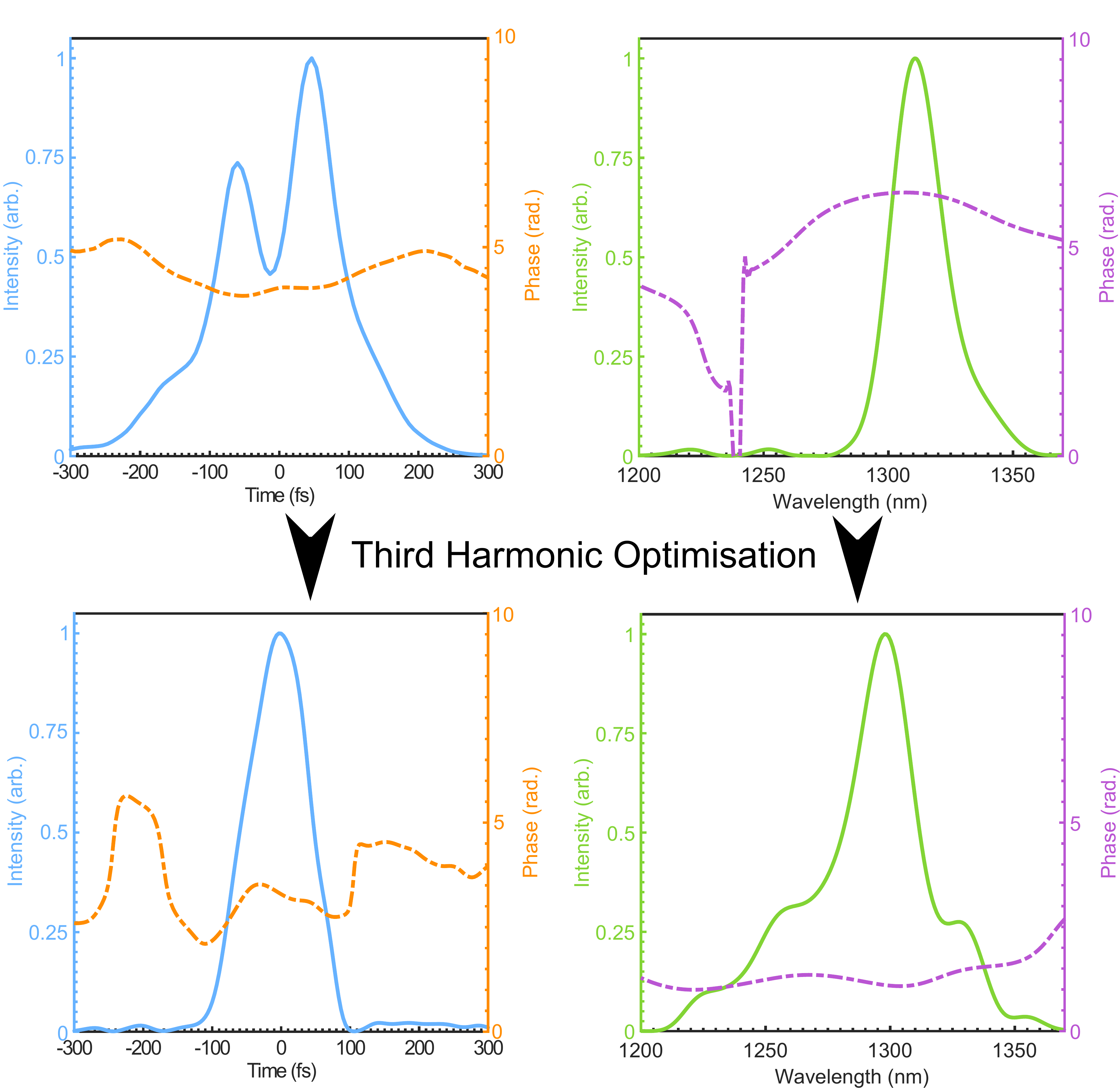}
   \end{tabular}
   \end{center}
   \caption[example] 
   { \label{Blind} 
Results of SHG FROG measurement on the OPA output for $\lambda=1300\:$nm. The left most plots show the temporal intensity (solid blue line) and phase (dashed orange line), while the right most show the spectral intensity (solid green line) and phase (dashed purple line). The upper plots are for when only the OPA pulse energy is optimised, while the lower plots are for when the third harmonic intensity, generated by focussing in air, is optimised.}
   \end{figure} 

In fig. \ref{Blind} we show the pulse duration of the OPA, measured using an SHG FROG, in two cases. The pulse shown in the upper panels was measured after the pulse energy of the OPA output was optimised only, whereas the lower panels shows the same pulse properties after the third harmonic was optimised instead. In the latter case the pulse is significantly shorter and less temporally structured after the ``blind'' optimisation procedure, while the measured pulse energy was just $~15\%$ lower, explaining the enhanced peak intensity. We also note that the brightest high-order harmonics coincided with the situation when the OPA was optimised using the third harmonic.
 
\section{HIGH HARMONIC GENERATION CALCULATION}
\label{Model}
In sections \ref{IRHHG} and \ref{VISHHG} we describe experiments where high-order harmonics were generated using both SWIR and VIS pulses from the OPA. To support this experimental work we have performed 1-D calculations of high harmonic generation in order to estimate the harmonic cut-off and phasematching pressure for different driver wavelengths and gas species.

For harmonic generation the wavevector mismatch can be written as $\Delta k = q k(\omega_0) - k (q \omega_0) + \Delta k_\mathrm{G}$, where $q$ is the harmonic order, $k(\omega)$ is the wavevector for radiation of angular frequency $\omega$, $\omega_0$ is the angular frequency of the fundamental beam, and $\Delta k_\mathrm{G}$ arises from geometric effects (e.g. the Gouy phase). If we assume that dispersion arises from neutral gas, plasma, and the Gouy phase, the phase-matching pressure -- i.e. the pressure for which $\Delta k = 0 $, given by \cite{Rothhardt2014b}:

\begin{equation}
P_\mathrm{m} = \frac{P_0 \lambda^2}{2 \pi^2 w_0^2 \Delta n(1-\frac{\eta}{\eta_\mathrm{c}})}
\label{Pm}
\end{equation}

where $P_0$ is the standard pressure, $\Delta n$ difference between the refractive index of the neutral gas at the wavelengths of the driver and harmonic, $w_0$ is the spot size, $\eta$ is the ionization fraction and $\eta_\mathrm{c}$ is the critical ionization fraction given by:
\begin{equation}
\eta_\mathrm{c} = \bigg(1+\frac{N_0 r_\mathrm{e} \lambda^2}{2\pi \Delta n}\bigg)^{-1}
\label{NoGouy}
\end{equation}
where $N_0$ is the number density at atmospheric pressure and $r_\mathrm{e}$ is the classical electron radius.

The ionization fraction will increase monotonically during the laser pulse and may be calculated as a function of time using the Yudin-Ivanov (YI) model.\cite{Yudin2001} The YI model accounts for both tunnelling and multiphoton contributions to the ionization fraction, and is valid over a broader range of Keldysh parameters than the more common ADK model.\cite{Ammosov1986} This is expected to be particularly relevant when using short wavelength drivers.

The cut-off law for high-harmonic generation can be written as:
\begin{equation}
\gamma_\mathrm{max}=I_\mathrm{p}+C I_\mathrm{eff}\lambda^2
\end{equation}
where $\gamma_\mathrm{max}$ is the highest generated photon energy, $I_\mathrm{p}$ is the ionization potential of the gas species, $C=9.33\times10^{-13}$\footnote{This value of C is valid where the units of $\gamma_\mathrm{max}$ and $I_\mathrm{p}$ are in eV, $\lambda$ is in microns and $I_\mathrm{eff}$ is in W/$\mathrm{cm}^2$}, $\lambda$ is the fundamental wavelength and $I_\mathrm{eff}$ is the intensity of the fundamental. The phasematching pressure remains finite whilst $\eta(t)<\eta_\mathrm{c}$; to estimate $P_\mathrm{m}$ we set $\eta(t)$ to its value at the moment during the laser pulse where the harmonic can be first generated according the cut-off law.

\section{HIGH ORDER HARMONICS GENERATED BY INFRARED PULSES}
\label{IRHHG}
With the final stage of the OPA set to difference-frequency mode, the centre wavelength was adjusted to $\lambda=1300\:\mathrm{nm}$. The peak intensity in situ (in the high harmonic generating vacuum chamber) was maximised as described in sec. \ref{BLIND} without additional dispersion compensation. The optimized pulses had an energy of $250\:\mu$J and an duration of $\approx100$ fs. The laser beam was directed into the vacuum chamber through a 1 mm thick anti-reflection coated window and then on to a spherical mirror of $f=15\:$cm at a near-normal angle of incidence. The gas cell comprised a thin-walled ($200\: \mu$m thick) hollow nickel tube, pressed to an outer diameter of 1.2 mm (comparable to the 1.5 mm Rayleigh range of the fundamental beam), with entrance and exit holes drilled by the focussed laser beam.

\begin{figure} [ht]
   \begin{center}
   \begin{tabular}{c} 
   \includegraphics[height=9cm]{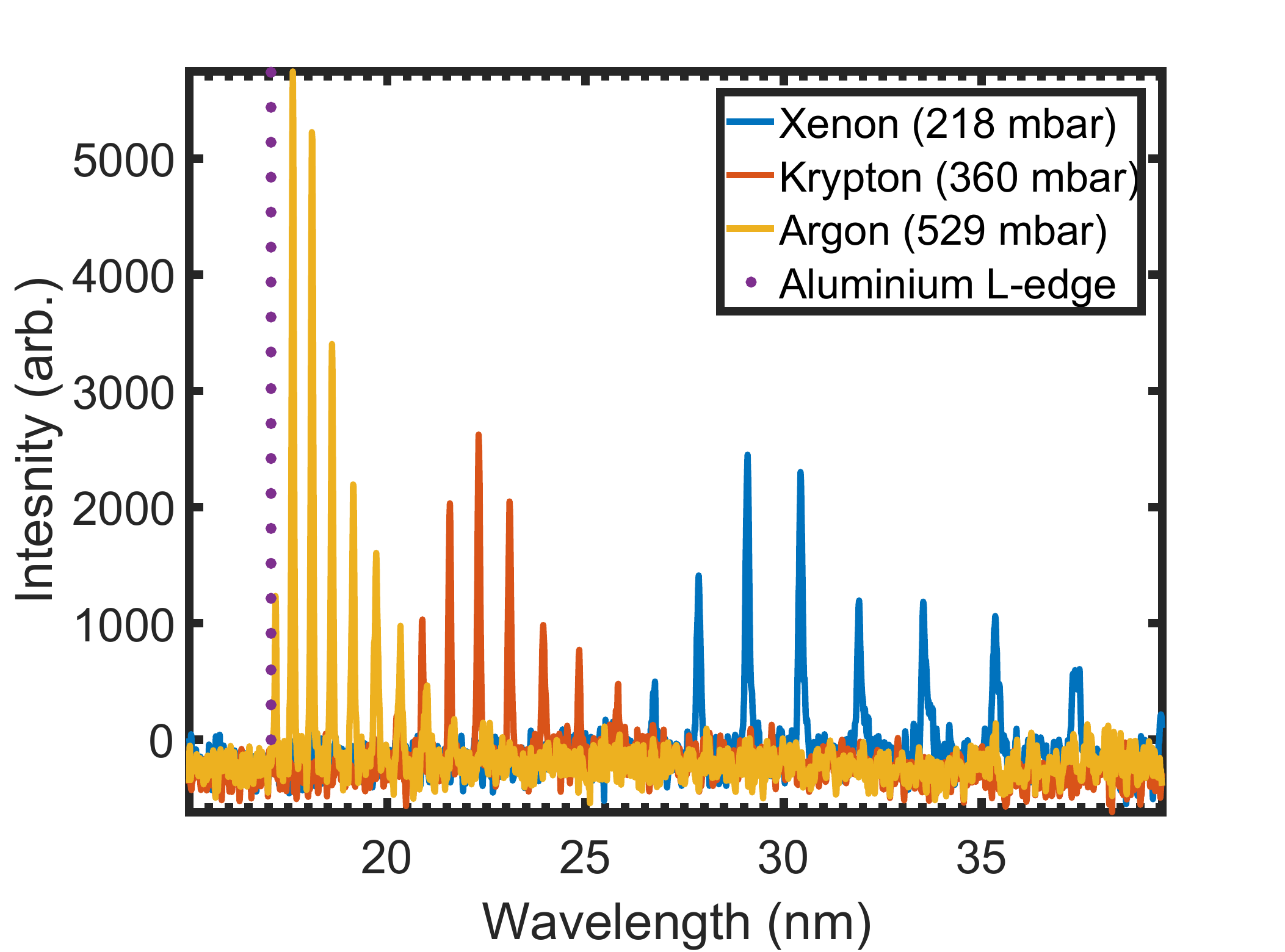}
   \end{tabular}
   \end{center}
   \caption[example] 
   { \label{IRHHG_} 
High harmonic spectra obtained for the same driver and focusing conditions but with the gas cell backed by 218 mbar of xenon (blue line), 360 mbar of krypton (red line) or 529 mbar of argon (yellow line). The location of the aluminium L-edge is marked by the purple dotted line.}
   \end{figure}

High-order harmonics were observed using a flat-field imaging spectrometer and x-ray CCD (Andor DO440-BN). For the experiments described here, unless otherwise stated, two 200 nm thick aluminium filters, supported by a fine nickel mesh, were used to separate the fundamental laser field from the high harmonic beam. Maintaining the same focussing conditions and OPA settings throughout, high harmonic generation in Xe, Kr and Ar was investigated, as shown in fig. \ref{IRHHG_}. The gas pressure which yielded the strongest signal on the CCD was found for each species, which we will take as the phasematching pressure. The harmonics range from 39.5 nm up to the aluminium L-edge at 17 nm. Subsequent tests with a 400 nm thick zirconium filter, instead of the Al filters, showed that the observable harmonic cut-off extended to 15.1 nm (the $87^\mathrm{th}$ harmonic order) when argon was the generating gas.

\begin{table}[ht]

\label{Comp_table}
\begin{center}       
\begin{tabular}{|l|l|l|l|l|}
\hline
\rule[-1ex]{0pt}{3.5ex}  Gas & $P_\mathrm{m}$ exp. (mbar) & $P_\mathrm{m}$ calc. (mbar)  & $\gamma_\mathrm{max}$ exp.  (nm) & $\gamma_\mathrm{max}$ calc. (nm) \\
\hline
\rule[-1ex]{0pt}{3.5ex}  Xenon & 220 & 218 & 26.7 & 23.8   \\
\hline
\rule[-1ex]{0pt}{3.5ex}  Krypton & 360 & 330 & 19.7 & 17.0  \\
\hline
\rule[-1ex]{0pt}{3.5ex}  Argon & 530 & 503 & 15.1 & 12.4  \\
\hline 
\end{tabular}
\end{center}
\caption{Comparison between calculated and experimentally determined phasematching pressures and harmonic cut-off energies for high harmonics generated in xenon, krypton and argon using $\lambda=1300 \:\mathrm{nm}$ pulses from the OPA. The phasematching pressure was evaluated for a harmonic photon energy of 35 eV. The fundamental was assumed to be a Gaussian pulse with an energy of $250\:\mu$J, a duration of $100\:$fs and spot size of $25\:\mu$m.} 
\end{table}

Table \ref{Comp_table} compares the measured values of $P_\mathrm{m}$ and $\gamma_\mathrm{max}$ with those calculated from the procedure outlined in sec. \ref{Model}. It can be seen that for both $P_\mathrm{m}$ and $\gamma_\mathrm{max}$ agreement between experiment and calculation is very good.

\section{HIGH ORDER HARMONICS GENERATED BY VISIBLE PULSES}
\label{VISHHG}

\begin{figure} [ht]
   \begin{center}
   \begin{tabular}{c} 
   \includegraphics[height=8cm]{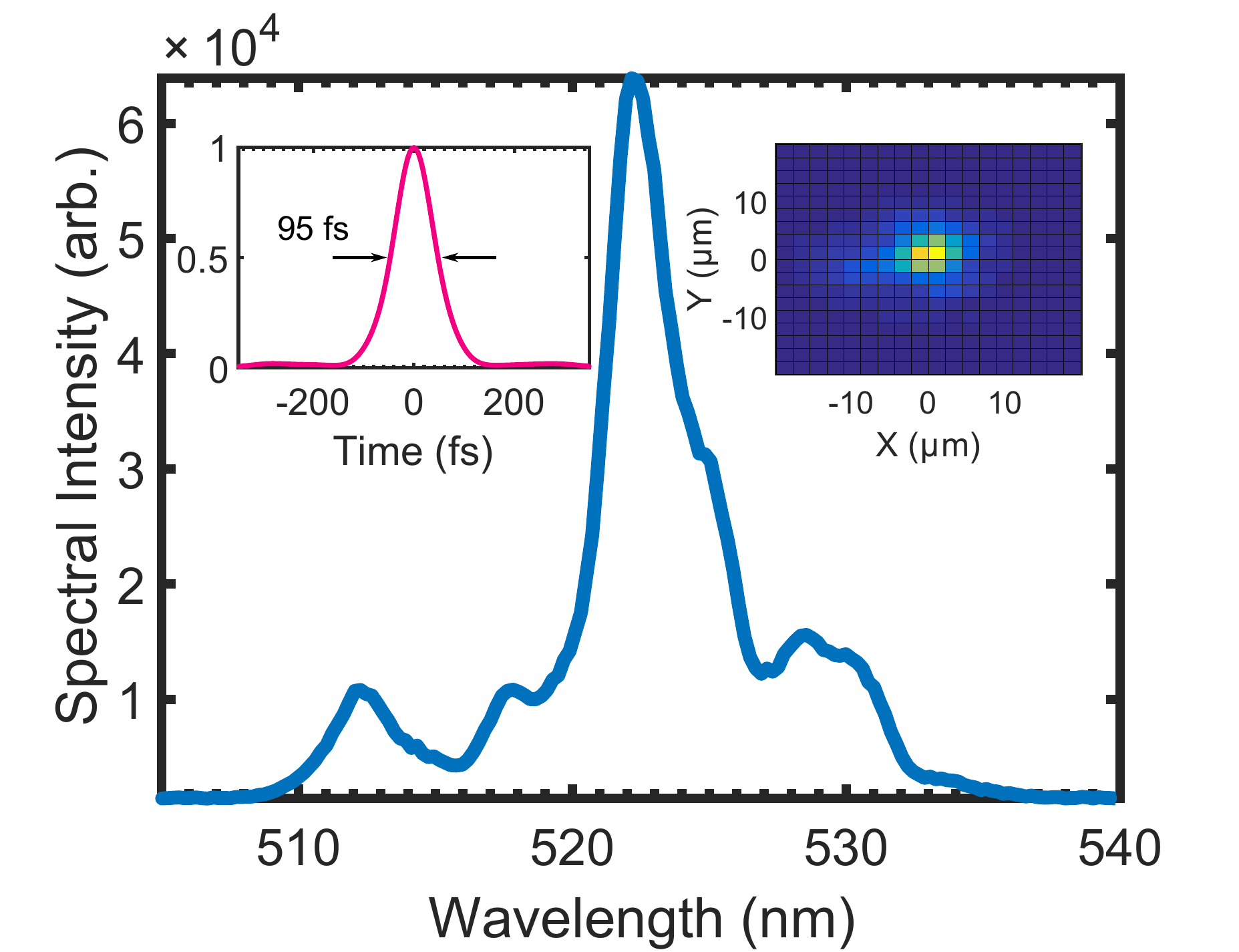}
   \end{tabular}
   \end{center}
   \caption[example] 
   { \label{GRNSPC} 
Main figure: Fundamental spectrum of the OPA output (in sum-frequency mode) used to generate high order harmonics. Top left inset: Corresponding Fourier-limited pulse temporal profile. Top right inset: OPA output focal spot recorded using a CCD.}
   \end{figure}

As a further demonstration of the flexibility of our OPA we generated high-order harmonics with visible laser pulses produced by the OPA in sum-frequency mode. The amplified signal beam from P1 was tuned to a wavelength of $\lambda_\mathrm{s} = 1502\:$nm, yielding an output wavelength of $\lambda=\:522\:$nm from the final stage. The output laser spectrum measured using a fibre coupled spectrometer (Ocean Optics USB4000) is shown in fig. \ref{GRNSPC}, corresponding to Fourier-limited duration of less than 100 fs (see inset of \ref{GRNSPC}). The pulse energy measured directly out of the OPA in this case was 172 $\mu$J. The beam was demagnified with a $M=0.5$ telescope, and directed into a vacuum chamber through an anti-reflection coated window. Owing to losses from an uncoated lens in the telescope, the pulse energy measured in the vacuum chamber was $130\:\mu$J. In the chamber the beam was tightly focused using a $F=75\:$mm achromatic lens. An image of the focal spot measured with a CCD is shown in the top-right inset of fig. \ref{GRNSPC}. A gas cell of a design similar to that described in section \ref{IRHHG} was located close to the focal plane. Owing to the smaller Rayleigh range in this case ($\approx 600\:\mu$m) a gas cell of outer diameter 700 $\mu$m was employed.

\begin{figure} [ht]
   \begin{center}
   \begin{tabular}{c} 
   \includegraphics[height=9cm]{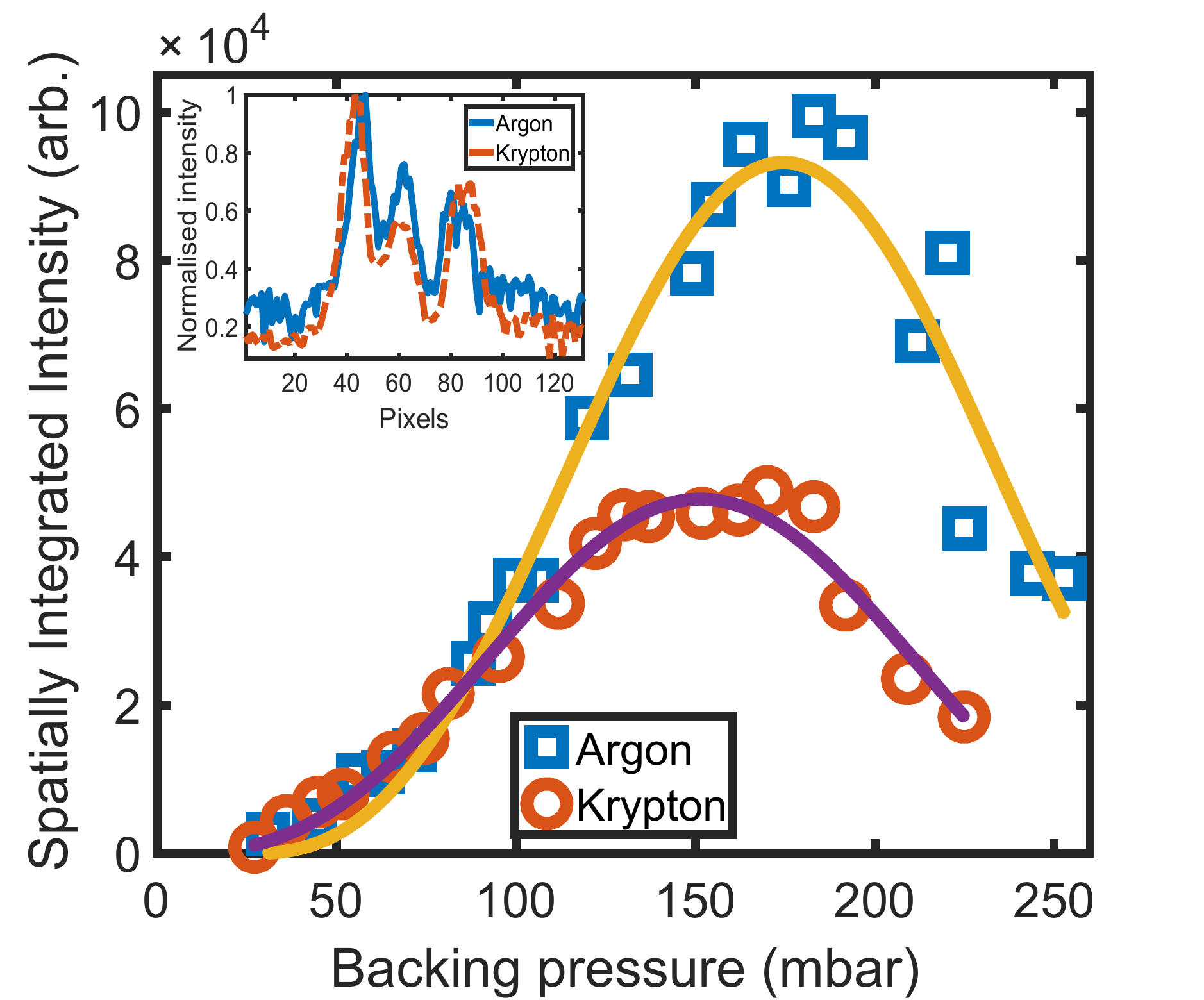}
   \end{tabular}
   \end{center}
   \caption[example] 
   { \label{GRNHHG} 
Main figure: spatially integrated high harmonic signal plotted as a function gas cell backing pressure for argon (blue squares) and krypton (red circles). A fit of the function $I= a\times \mathrm{sinc}(bP+c )^2$ is shown for both argon (solid yellow line) and krypton (solid purple line). Inset: Recorded fringe pattern in the case that the gas cell was backed with 185 mbar of argon (solid blue line) or 142 mbar of krypton (red dashed line).}
   \end{figure}

Using the model from section \ref{Model} and the experimental laser parameters, we calculated the harmonic cut-off to be 56.3 nm in the case of krypton and 44.6 nm for argon. The model also gives $I_\mathrm{eff}$ as $3.14\times10^{13}\:\mathrm{W cm}^{-2}$ and $4.74\times10^{13}\:\mathrm{W cm}^{-2}$ for krypton and argon respectively, equivalent to Keldysh parameters of 2.96 and 2.56, i.e. in the multiphoton ionization regime. The wavelengths of the expected harmonic cut-offs were too long to be dispersed by our spectrometer, so instead the harmonics were detected in zero-order.

In lieu of a direct measurement of the harmonic spectrum, we record interference patterns produced by illuminating a slit pair (centre-to-centre separation of $100\:\mu$m, $50\:\mu$m slit width) with the high harmonic beam. The results are shown in the inset of fig. \ref{GRNHHG}. It should be noted that the fringe patterns exhibit additional modulations caused by the nickel mesh of the aluminium filters. Previously, two-slit interference patterns have been used to infer the spectrum of high harmonic beams \cite{Dilanian2008}. While we do not apply a similar analysis here, we note that the spacing of the fringes obtained with Kr is broader than those obtained with Ar, which is consistent with the longer wavelength harmonic cut-off estimated for the former gas species.

Plots of the (2-D) spatially integrated harmonic signal as a function of pressure are shown for both gases in fig. \ref{GRNHHG}. The function $I= a\times \mathrm{sinc}(bP+c)^2$ where $I$ is the spatially integrated harmonic intensity and $P$ is the gas cell backing pressure, was fitted to the two datasets with $a$, $b$ and $c$ as fit parameters. The fit is derived from the harmonic intensity dependence on wavevector mismatch, with $\Delta k\propto P$ as per equation \ref{Pm}. The fit yields a phasematching pressure for krypton of 151 mbar, while for argon the phasematching pressure was 174 mbar. Additionally the maximum harmonic intensity measured for argon was twice that obtained for krypton. For laser parameters of $\lambda = 522\:$nm, $E = 130\:\mu$J, $\tau = 150\:$fs and $w_0 = 14\:\mu$m, and a photon energy of 16.8 eV ($q=7$), the calculation for the high harmonic phasematching pressure yields $P_\mathrm{m} = 167\:$mbar for krypton and $P_\mathrm{m} = 250\:$mbar for argon, in reasonable agreement with the experimentally derived values.     

\section{FUTURE APPLICATIONS}
\label{Apps}

\begin{figure} [ht]
   \begin{center}
   \begin{tabular}{c} 
   \includegraphics[height=10cm]{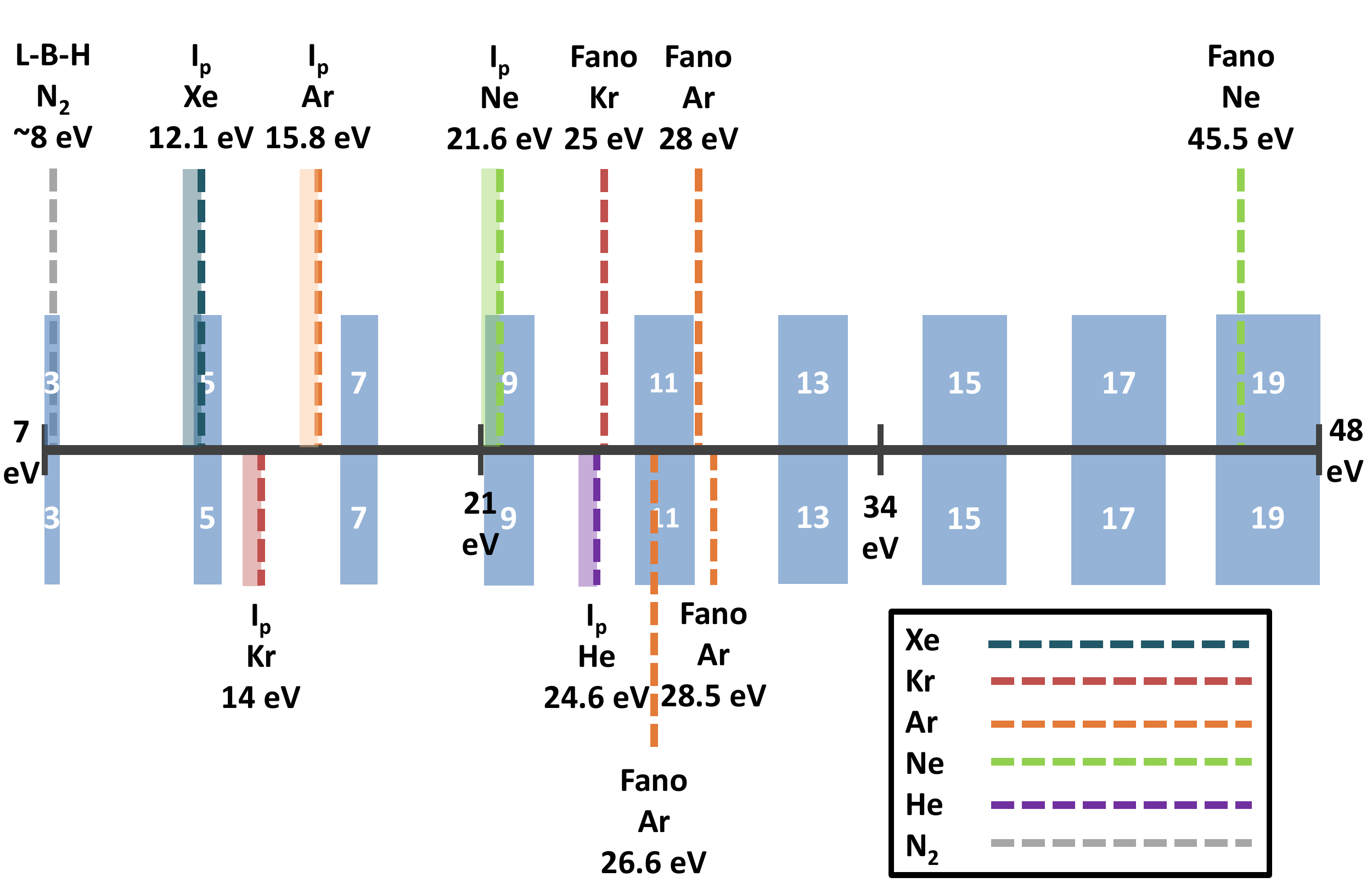}
   \end{tabular}
   \end{center}
   \caption[example] 
   { \label{eHHG} 
Spectral coverage of the high harmonics produced by visible wavelength pulses from the OPA in the region 7--48 eV. The harmonic orders are labelled with white numbers. Overlaid are the ionization potential locations and selected atomic resonances in various noble gases. The location of the three photon resonant Lyman-Birge-Hopfield (L-B-H) manifold in $N_2$ is also shown.}
   \end{figure} 

The results presented in sections \ref{IRHHG} and \ref{VISHHG} establish that it is possible to use our OPA to generate high-order harmonics over a wide span of driver wavelengths. Indeed, the combination of GW level peak power and tunable, visible wavelength allow for a number of interesting phenomena to be studied. In this section, for the sake of brevity, we limit the discussion to the possibility of experimentally studying resonantly excited atomic and molecular states in the context of HHG using a similar laser source as described in this paper.

To this end, we show in fig. \ref{eHHG} the photon energies covered by high harmonics driven by visible pulses from our OPA in the range 7--48 eV. While the wavelength tuning range we achieve may seem small ($<\lambda/10$), it is large enough to allow access to a number of atomic resonances, marked in fig. \ref{eHHG}. For instance, the Fano-type resonances in argon between 26.6 and 28.5 eV. One such state coincides with the 11th harmonic order (specifically when the centre wavelength is tuned to $\approx 513\:$nm). Such a resonance has been previously exploited to produce a bright, narrow-band, spectrally isolated harmonic beam \cite{Rothhardt2014}, well-suited to applications such as angle-resolved photoelectron spectroscopy and coherent diffractive imaging. This arrangement would also benefit from the favourable wavelength scaling of the single atom response to the fundamental field for visible pulses, compared to longer wavelength driver pulses.

Further, the ionization potential in neon is coincident with the 9th harmonic order, while the ionization potential in xenon is coincident with the 5th order. In both cases it could be possible to observe free-induction decay at photon energies just below the ionization potential level \cite{Bengtsson2016}. Free-induction decay occurs from the population of an excited state, which subsequently decays over relatively long period of time ($\approx$ ps), producing narrow-band emission, and has previously been observed in argon \cite{Bengtsson2016}. Utilising the free induction decay emission has been mooted as a novel light source for applications requiring narrowband XUV light, such as precision spectroscopy.

Finally, the Lyman-Birge-Hopfield (LBH) manifold in molecular nitrogen is marked in fig. \ref{eHHG} around 7-8 eV, coincident with the third harmonic. Previously Dousott and co-workers exploited a three-photon transition in krypton to stabilise and extend a laser induced filament at 300 nm through the high-order Kerr effect \cite{Doussot2016}. It could be possible to use the tunable output of our OPA with $\mathrm{N}_2$ to populate LBH states before observing high harmonic generation from the excited molecular configuration. Such a process, recently demonstrated in argon by Beaulieu \emph{et al.} \cite{Beaulieu2016} has been dubbed excited state high harmonic generation or e-HHG, and represents an intriguing development in strong field physics. The extension of these new techniques to molecules could be a useful tool for probing the behaviour of atoms excited by strong, resonant laser fields. 

\section{Conclusion}

We have constructed an OPA capable of producing pulses with GW-level peak power in both the shortwave infrared and visible spectrum. The OPA has a reconfigurable final stage, capable of either parametric amplification of a SWIR seed beam or sum-frequency generation between the seed and the pump. This approach has the possibility of widening the spectral coverage available from extant OPAs.

The SWIR pulses produced by the OPA are $\approx 100\:$fs in duration and are close to the transform-limit in part due to the low net dispersion of the OPA design. The visible pulses have a more structured temporal profile, compared the SWIR, yet contain most of the pulse energy within $\approx120\:$fs. For $3\:$mJ of input pump pulse energy, SWIR pulses with an energy of $>350\:\mu$J or visible pulses with an energy of $192\:\mu$J were produced.

Both visible and SWIR pulses were used to generate high-order harmonics from various gas species. For an fundamental wavelength of $\lambda=1300\:$nm harmonics up to the $87^\mathrm{th}$ order (15.1 nm wavelength) were produced in argon. The high-order harmonics generated by the visible pulses were detected in zero-order by a spectrometer. In all cases, the measured phasematching pressure was found to agree closely with the results of a simple, 1-D model of the harmonic generation process. 

We anticipate that sources of tunable, visible wavelength pulses, like that described here, will find use in the development of recently demonstrated techniques exploiting atomic and molecular resonances in various gas species.

\newpage
\acknowledgments 
 
The authors would like to thank Vikaran Khanna and Jens Biegert for helpful advice in designing the OPA. The authors would also like to thank Ian Walmsley for fruitful discussions. This work was supported by EPSRC (grant number EP/L015137/1). 

\bibliography{Photonics_West_references} 
\bibliographystyle{spiebib} 

\end{document}